\documentclass[aps,prl,twocolumn,showpacs,amsmath,amssymb]{revtex4}

\begin{document}
\bibliographystyle{revtex}
\title{The analytical solution of the Schr\"{o}dinger equation in
Born-Oppenheimer approximation for $H_2^+$ molecular ion}

\vspace{0.5cm}
\author{Alexander V. Mitin} \email{mitin@diku.dk}
\affiliation{Datalogisk Institut, University of Copenhagen,
Universitetsparken 1, 2100 Copenhagen, Denmark}

\date{\today}
\begin{abstract}
An analysis of the analytical solution of the Schr\"{o}dinger
equation (which is a second order differential equation) for
$H_2^+$ shows that the second linear independent solution of this
equation is a square integrable function and therefore the ground
state total wave function is a linear combination of two linear
independent wave functions of different space symmetry:
cylindrical and spherical. The wave function of cylindrical
symmetry is well known. It has maxima at the positions of nuclei.
The wave function of spherical symmetry and the corresponding
spherical electron distribution, which exists at $R\neq0$ and
locates at the middle of the bond, represents a quasiatom of
electron density of non-nuclear united atom. In the light of the
new result the qualitative behavior of the ground state wave
function and the electron density of $H_2^+$ has been
reinvestigated. It is shown analytically that a transformation of
the total molecular wave function with two maxima to that one with
one maximum passes through a flat wave function. The presented
three-dimension figures of the electron density visualize the
spherical component of the total wave function and its
transformation with increasing internuclear separation.
\end{abstract}
\pacs{31.10.+z, 03.65.-w}
\maketitle

\section{}

The property and behavior of the ground state wave function of
$H_2^+$ molecular ion is very important for quantum mechanics
because it establishes our general representation on the quantum
interaction in molecules. For this reason the ground state wave
function and electron density of $H_2^+$ ion was investigated in
many publications by using different approaches. However, an
analysis of these results shows that there is a direct
contradiction between them. This lead us to non-unique qualitative
representation on a behavior of the ground state wave function.

Thus, from one hand side, from asymptotic properties of the total
wave function follow that at the limit of a united atom the ground
state wave function of $H_2^+$ transforms to the wave function of
$He^+$. Therefore, after some small $R$ the $H_2^+$ wave function
must have only one maximum at the middle of the bond. This is a
basis of perturbation theory developed by Bethe for a calculation
of the $H_2^+$ total energy at short $R$ \cite{Bethe}. In this
theory the $H_2^+$ wave function is approximated by the $He^+$
wave function located at the middle of the bond. The investigation
of Bethe's perturbation theory shows that it works well up to
$R=0.05$ $a.u.$ \cite{Wind}. Other example, which displays an
importance of a correct asymptotic of the total molecular wave
function in the limit of a united atom was given in Ref.
\cite{Dalgarno}. In this work it was shown that a representation
of the molecular wave function as a linear combination of united
and separated atoms wave functions results in significant
improvement of the $H_2^+$ variational total energy calculated at
short $R$.

From other hand side, from a solution of the Schr\"{o}dinger
equation in Born-Oppenheimer approximation in alliptic coordinates
for $H_2^+$ given by Burrau \cite{Burrau} follows that the
solution has two maxima at the position of two nuclei.

To clarify this question and to show that the second linear
independent solution of the Schr\"{o}dinger equation is square
integrable function let us consider first the $H_2^+$ problem
qualitatively. In the Born-Oppenheimer approximation the
Hamiltonian of $H_2^+$ can be written as
\begin{eqnarray}
H(R)=-\frac{\hbar^2}{2m}{\nabla^2} - \frac{1}{\mid r-R_A
\mid}-\frac{1}{\mid r-R_B \mid} + \frac{1}{R} \label{eq1}
\end{eqnarray}
where $r$ is the vector of the electron; $R=\mid R_A -R_B \mid$ is
the internuclear distance; $R_A$ and $R_B$ are the vectors of
nuclei $A$ and $B$; and the electron and nuclei charges are equal
to one. A parametric dependence of $H(R)$ on $R$ shows that at the
limit $R\to0$ Hamiltonian (\ref{eq1}) reduces to the Hamiltonian
of the $He^+$
$$
H(R)=-\frac{\hbar^2}{2m}{\nabla^2}-\frac{2}{\mid r-R_A\mid} ,
$$
while the $H_2^+$ total wave function transforms to the total wave
function of the $He^+$ ion. At the limit $R\to\infty$ there is an
equally probability that the electron can be located near nucleus
$A$ or $B$ and thus forms $H$ atom at the point $A$ or $B$.
Therefore, Hamiltonian (\ref{eq1}) transforms to the Hamiltonian
of the $H$ atom
$$
H(R)=-\frac{\hbar^2}{2m}{\nabla^2} - \frac{1}{\mid r-R_A \mid} .
$$
The total molecular wave function in this case reduces to the wave
function of the $H$ atom.

Summarize the asymptotics of Hamiltonian (\ref{eq1}) and follow to
the {\it superposition principle} we can conclude that the total
wave function of $H_2^+$ molecular ion can be presented as a
linear combination of three atomic wave functions
\begin{equation}
\label{eq2} \Psi(R)=c_A(R)\Psi_A+c_B(R)\Psi_B+c_U(R)\Psi_U
\end{equation}
where $\Psi_A$ and $\Psi_B$ are the wave functions of $H$ atoms
and $\Psi_U$ is the total wave function of $He^+$ ion. In this
expression each wave function $\Psi_A$, $\Psi_B$, and $\Psi_U$ is
located on a separated center. Therefore, we can see that
consistent quantum mechanical consideration of $H_2^+$ problem
results in a three-center problem with corresponding three-center
representation of the total molecular wave function.

The use of representation ({\ref{eq2}}) for the total molecular
wave function in calculations of $H_2^+$ total energy with high
precision at short $R$ was advocated in Ref. \cite{Dalgarno}. At
equilibrium distances the use two-center expansion instead
three-center expansion ({\ref{eq2}}) results in a small gap
between the Hartree-Fock total energies of $H_2$ calculated by
numerical methods and in conventional two-center LCAO
approximation. In Ref. \cite{Mitin} was shown that the use of
expansion ({\ref{eq2}}) permits to remove this gap. These two
examples together with the Bethe's perturbation theory
\cite{Bethe} show that the expansion ({\ref{eq2}}) is correct.

One of a particular interest of the present study consists in
study of a transformation of the molecular wave function and
electron density distribution with changing $R$ from $R=0$ to
$R=+\infty$. Expression (\ref{eq2}) together with a conventional
assumption that a wave function and its derivatives are continuous
functions gives us a solid basis for a consideration of such
transformation.

From the consideration given above follows that if $\Psi_A$,
$\Psi_B$, and $\Psi_U$ are atomic wave functions then at $R=0$
$c_A(0)=c_B(0)=0$ and $c_U(0)=1$, while at $R=+\infty$
$c_A(+\infty)=c_B(+\infty)=c_N$, where $c_N$ is a normalization
coefficient and $c_U(+\infty)=0$. Now, taking into account a
continuous property of $\Psi(R)$, which means that all
coefficients in (\ref{eq2}) and their derivatives are continuous
functions on $R$, and suppose that a distribution of electron
density is defined by a total wave function, we can derive two
important statements:

a) for any small $\varepsilon_0$ it is possible to find such $R_0$
that when $R<R_0$ the total molecular wave function can be
presented as $\Psi(R)=c_U(R)\Psi_U+O(\varepsilon_0)$. This means
that at sufficient small $R$ the total molecular wave function and
corresponding distribution of electron density must have only one
maximum. As a consequence, a transformation of the molecular wave
function and the electron density with two maxima to those with
one maximum passes through a flat wave function and corresponding
flat electron distribution.

b) the spherical component $\Psi_U$ of the total molecular wave
function $\Psi(R)$ exists in molecule at $R\neq0$. It is the wave
function of a united atom and in molecule it represents the
electron density of a non-nuclear united atom or quasiatom. The
spherical component of the molecular wave function and appropriate
electron distribution located at the middle of the bond can be
observed in molecule at some $R$.

Now, follow to Burrau \cite{Burrau}, we consider the
Schr\"{o}dinger equation in Born-Oppenheimer approximation for
$H_2^+$ in elliptic coordinates: $\xi=\left(r_1+r_2\right)/R$ and
$\eta=\left(r_1-r_2\right)/R$. In these coordinates, the wave
function for the $\Sigma$ ground state can be presented as
$\Psi=X(\xi,R)Y(\eta,R)$, where $X(\xi,R)$ and $Y(\eta,R)$
satisfies the following equations:
$$
\left(\xi^2-1\right)\frac{d^2X}{d\xi^2}+2\xi\frac{dX}{d\xi}+
\left(\frac{1}{2}ER^2\xi^2-2R\xi+A\right)X=0
$$
\begin{eqnarray}
\left(1-\eta^2\right)\frac{d^2Y}{d\eta^2}-2\eta\frac{dY}{d\eta}-
\left(\frac{1}{2}ER^2\eta^2+A\right)Y=0 \label{eq3}
\end{eqnarray}
and $E$ and $A$ is energy and separation constant to be
determined. It needs to note here that the elliptic coordinates
are undefined at $R=0$ and therefore the equations given above are
valid only for $R>0$.

To clarify behavior of the total wave function at the middle point
it is sufficient to investigate behavior of $Y\left(\eta,R\right)$
and $Y'\left(\eta,R\right)$ at $\eta=0$ and at small $R$. In this
case equation (\ref{eq3}) transforms to an ordinary differential
equation of the second order
$$
\frac{d^2Y}{d\eta^2}-AY=0
$$
The general solution of this equation is as a linear combination
of two linear independent solutions:
\begin{eqnarray}
Y(0,R)=c_1\left(R\right)\exp{\left(-\sqrt{A}\vert\eta\vert\right)}+
c_2\left(R\right)\exp{\left(\sqrt{A}\vert\eta\vert\right)}
\label{eq4}
\end{eqnarray}
that must be taken at $\eta=0$. Despite of this limitation we can
conclude that the algebraic structure of the total wave function
will be preserve at the vicinity of this point because of the wave
function is continuous function. This permits us to compare the
expressions (\ref{eq4}) and (\ref{eq2}). From this comparison
follows that the first solution of spherical symmetry corresponds
to the wave function of a united atom which has a maximum on
$\eta$ at $\eta=0$, while the second one of cylindrical symmetry
has {\it no local minimum} on $\eta$ at $\eta=0$ and is formed by
the linear combination of wave functions of separated atoms. On
this basis we can conclude that the both linear independent
solutions are square integrable functions. Therefore, the general
solution of the Schr\"{o}dinger equation indeed is a linear
combination of two linear independent solutions that represent two
different quantum objects.

It is interesting to note that Eq. (\ref{eq4}) is similar to the
one-dimension Schr\"{o}dinger equation for an electron in a
central potential
$$
\frac{d^2\Psi}{dx^2}+(V-E)\Psi=0
$$
A formal solution of this equation is also a linear combination of
two linear independent solutions
$$
\Psi=d_1\exp{\left(-\sqrt{E-V}\vert x\vert\right)}+
d_2\exp{\left(\sqrt{E-V}\vert x\vert\right)}
$$
However, in this case the second solution is not square integrable
function. Therefore, it is excluded from a considered.

Now, suppose that the coefficients in (\ref{eq4}) are continuous
functions, we can conclude that at $R\to0$ Eq. (\ref{eq4}) can be
smoothly transformed to the well-known one-dimensional atomic wave
function only if $c_2(R)\to0$ (because of this component
transforms to the not square integrable function) and $c_1(R)$
becomes equal to a normalization constant of the united atom wave
function.

The first derivative of (\ref{eq4}) on $\eta$ taken at $\eta=0$ is
$$
\frac{dY(0,R)}{d\eta}=-\sqrt{A}\left[c_1\left(R\right)-
c_2\left(R\right)\right] .
$$
It is negative at small $R$  because of $c_2(R)\to0$ at $R\to0$.
This means that $Y\left(0,R\right)$ has a maximum. With increasing
$R$ and growing $c_2(R)$ the first derivative becomes equal to
zero and after that it can be positive or negative in dependence
on the values of the coefficients. At large $R$ when coefficient
$c_2(R)$ becomes dominant only "minimum" can be observed at
$\eta=0$.

In the light of the analytical solution of the Schr\"{o}dinger
equation given above it needs to comment the results presented in
\cite{Kato,Hoffmann}. Our analysis will base on the fact that the
Schr\"{o}dinger equation is a second order differential equation
and its general solution is a linear combination of two linear
independent solutions. An additional requirement for this equation
is that the solutions must be square integrable functions.
Otherwise they should be excluded from a consideration.

An analysis of the publications \cite{Kato,Hoffmann} shows that in
these studies only one solution of cylindrical symmetry (that can
be called classical) of the Schr\"{o}dinger equation has been
taken into consideration, while the existence of the second
solution and its analytical properties was not discussed. This
means that an {\it implicit assumption} that the second solution
is a not square integrable function has been used in these works.
However, the present study gives strong evidences that the second
solution of spherical symmetry (that can be called quantum or
quasiatomic) is a square integrable function and therefore it must
be taken into consideration also.

A proof that the results presented in \cite{Kato,Hoffmann} have
been obtained only for the classical component of the total wave
function but not for the total wave function follows directly from
the fact that the statement (v) of $Theorem~1$ \cite{Hoffmann}
that the wave function has no local minimum is a simple
consequence of the analytical solution given above.

To visualize the transformation of the total wave function
$\Psi(R)$, the electron densities of $H_2^+$ at different $R$ have
been calculated in quantum mechanical variational calculations by
using the Gaussian~98 program \cite{g98}. These calculations can
be considered also as an independent verification of the
analytical results presented above because only the nuclear
centered basis of 15{\it s}6{\it p}5{\it d}4{\it f}3{\it g}2{\it
h}1{\it i} Gaussian spherical functions was employed. It was
formed from an optimized set of 15{\it s} functions supplied by
polarization functions taken from pV7Z basis for $H$ atom
\cite{Feller}. A quality of the variational wave function was
controlled by comparing obtained variational total energies with
those ones calculated by the program \cite{Sattin} and presented
in Table I. A uniform deviation of the variational total energies
from the numerical ones, which is equal to $1.2*10^{-7}$ $E_h$,
points out that the employed basis results in a high quality
variational wave function.

The calculated electron densities of $H_2^+$ at a few $R$ are
presented on Fig.~1. At the beginning of a consideration we can
note that at $R=0$ the electron density is spherical. Then, at
small $R$, the spherical distribution continuously transforms to
an ellipsoidal distribution with one maximum. This distribution is
presented on Fig.~1 at $R=0.008$~$a.u.$ Further increasing $R$
results in appearing a flat surface on the electron density
distribution at $R=0.010$~$a.u.$ given on Fig.~1 followed by
appearing the two maxima at the positions of two nuclei and a
transformation of the ellipsoidal electron density distribution to
a cylindrical one. Appropriate cylindrical electron density at
$R=0.012$~$a.u.$ is presented at Fig.~1. Subsequent increasing $R$
reveals the existence of the spherical component of electron
density with additional maximum located at the middle of the bond.
The three-dimension figure of the electron density with a
quasiatom is presented on Fig.~1 at $R=0.019$~$a.u$.

To show that a quasiatomic solutions exist in any molecular system
the electron density distribution has been investigated in all
homonuclear diatomic molecules of the first-row elements. The
spherical symmetry electron density distributions (quasiatoms)
arisen from the second linear independent solutions and located at
the middle points have been observed in all molecules. At some
internuclear separations quasiatomic solutions result in
additional maxima on a profile of electron densities along the
molecular axis. An example of such electron density distribution
in $Li_2$ at 5.0 a.u. ($R_e$=5.051 a.u. \cite{HH79}) is presented
on Fig.~2. The quasiatom located at the middle point is explicitly
recognizable on this figure. The electron density distribution was
obtained in {\it ab initio} calculations with cc-pVQZ Gaussian
basis set \cite{qz} by quadratic configuration interaction method
as realized in Gaussian~98 program. Details of this study will be
given elsewhere \cite{Mitin2}.

The qualitative behavior of electron densities presented on these
figures is fully consistent with that one, which follows from the
analytical solution of equation (\ref{eq4}) and a qualitative
consideration of the asymptotic properties of the Hamiltonian (1)
given above. Despite of using only nuclear centered basis, that
corresponds to the cylindrical solution, the spherical component
of the molecular wave function or the other linear independent
solution of the Schr\"{o}dinger equation and the corresponding
spherical electron distribution located at the middle of the bond
was formed in accordance with theory of the second order
differential equation to keep asymptotic and continuous properties
of the total molecular wave function.

Thus, the analytical solution of the Schr\"{o}dinger equation
given in the present investigation shows that the ground state
total wave function of $H_2^+$ is a linear combination of two
linear independent wave functions of spherical and cylindrical
symmetry. The former one represents a quasiatom of the electron
density of non-nuclear united atom and is responsible for
appearing a spherical distribution of the electron density and an
additional maximum on a profile of the electron density along the
internuclear axis at some $R$.

\newpage

\begin{table}
\caption{\label{tab:table1} The total energies (in Hartree) of the
$H_2^+$ molecular ion.}

\begin{ruledtabular}
\begin{tabular}{ccc}
 $R$ (a.u.) & $E_{tot}$ (var.) &  $E_{tot}$ (numer.$^*$) \\
\hline 0.008 & 123.00016807 & 123.00016795 \\
0.010 & 98.00026148 &
98.00026136
\\ 0.012 & 81.33370830 & 81.33370818 \\
0.019 & 50.63250557 & 50.63250545
\\

\end{tabular}
\end{ruledtabular}

$^*$ - Calculated by the program \cite{Sattin}.
\end{table}

\mbox{}

\newpage

\vspace{2cm}

{\bf \large FIGURE CAPTIONS}

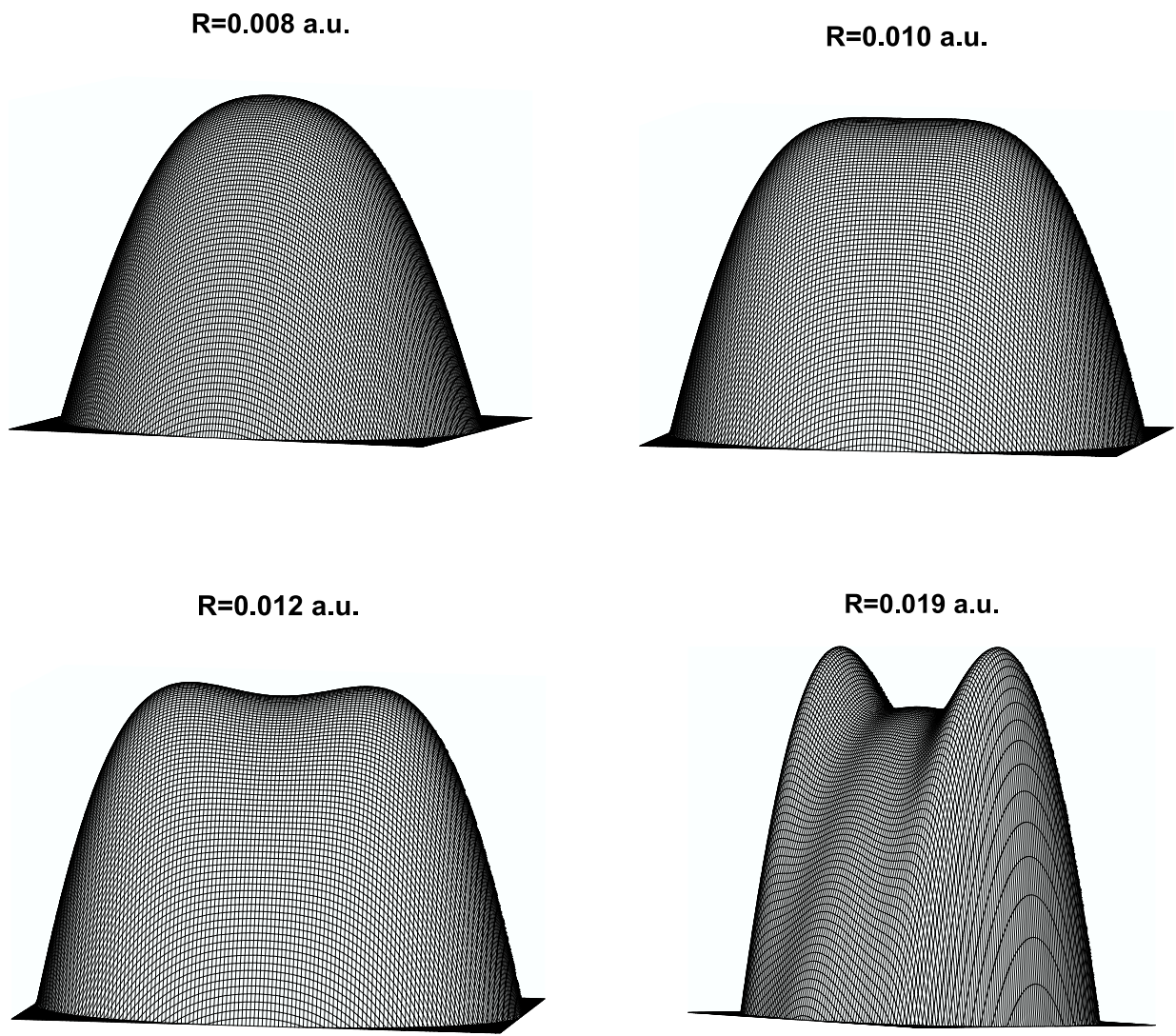
\vspace{0.5cm} \indent {\bf Figure 1}: The total electron
densities of $H_2^+$ molecular ion at different internuclear
distances.

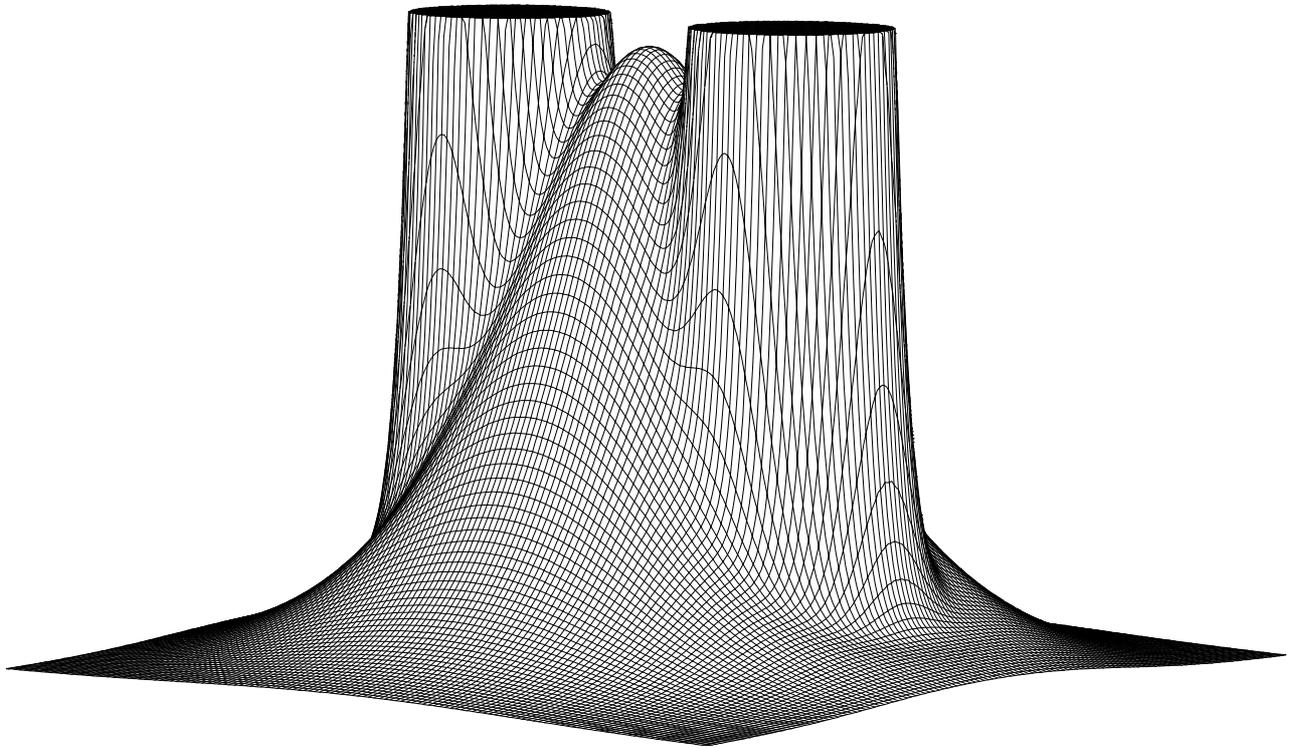
\vspace{0.5cm} \indent {\bf Figure 2}: The total electron
densities of $Li_2$ at 5.6~a.u.

\begin{thebibliography}{99}
\bibitem{Bethe}
    H. Bethe, in {\it Handbuch der Physik} Eds. H. Geiger and K. Scheel (Springer-Verlag, Berlin,
    1933), Vol. 24/1, p. 527.
\bibitem{Wind}
    H. Wind, J. Chem. Phys. {\bf 42}, 2371 (1965).
\bibitem{Dalgarno}
    A. Dalgarno and G. Poots, Proc. Phys. Soc. (London) {\bf 67A}, 343 (1954).
\bibitem{Burrau}
    \O. Burrau, Kgl. Danske Videnskab. Selskab. Math.-fysiske Meddel. {\bf 7}, No. 14, 1 (1927).
\bibitem{Mitin}
    A. V. Mitin, Phys. Rev. A {\bf 62}, 010501(R) (2000).
\bibitem{Kato}
    T. Kato, Commnun. Pure Appl. Math. {\bf 105}, 151 (1957).
\bibitem{Hoffmann}
    T. Hoffmann-Ostenhof and J. D. Morgan III, J. Chem. Phys. {\bf 75}, 843 (1981).
\bibitem{g98}
    Gaussian 98, Revision A.11.3, M. J. Frisch, G. W. Trucks, H. B. Schlegel,
    G. E. Scuseria, M. A. Robb, J. R. Cheeseman, V. G. Zakrzewski, J. A. Montgomery, Jr.,
    R. E. Stratmann, J. C. Burant, S. Dapprich, J. M. Millam, A. D. Daniels, K. N. Kudin,
    M. C. Strain, O. Farkas, J. Tomasi, V. Barone, M. Cossi, R. Cammi, B. Mennucci,
    C. Pomelli, C. Adamo, S. Clifford, J. Ochterski, G. A. Petersson, P. Y. Ayala, Q. Cui,
    K. Morokuma, N. Rega, P. Salvador, J. J. Dannenberg, D. K. Malick, A. D. Rabuck,
    K. Raghavachari, J. B. Foresman, J. Cioslowski, J. V. Ortiz, A. G. Baboul,
    B. B. Stefanov, G. Liu, A. Liashenko, P. Piskorz, I. Komaromi, R. Gomperts,
    R. L. Martin, D. J. Fox, T. Keith, M. A. Al-Laham, C. Y. Peng, A. Nanayakkara,
    M. Challacombe, P. M. W. Gill, B. Johnson, W. Chen, M. W. Wong, J. L. Andres,
    C. Gonzalez, M. Head-Gordon, E. S. Replogle, and J. A. Pople, Gaussian, Inc.,
    Pittsburgh PA, 2002.
\bibitem{Feller}
    D. Feller and K. A. Peterson, J. Chem. Phys. {\bf 110}, 8384 (1999).
\bibitem{Sattin}
    F. Sattin, Comp. Phys. Comm. {\bf 105}, 225 (1997).
\bibitem{HH79}
    K. P. Huber and G. Herzberg, {\it Constants of Diatomic Molecules},
    van Nostrand-Reinhold (1979).
\bibitem{qz}
    T. H. Dunning, Jr., J. Chem. Phys. {\bf 90}, 1007 (1989).
\bibitem{Mitin2}
    A. V. Mitin, to be published.
\end{thebibliography}
\end{document}